\documentclass[reprint,twocolumn,superscriptaddress,showkeys,
nofootinbib,notitlepage,amsmath,amssymb,floatfix]{revtex4-2}

\pdfoutput=1
\usepackage{latexsym,amsmath,amssymb,lmodern,float,url}
\usepackage{natbib}
\usepackage{appendix}
\usepackage{comment}

\usepackage{mathtools}
\usepackage{xcolor}

\def\be{\begin{equation}}
\def\ee{\end{equation}}
\def\M{\mathcal{M}}

\begin{document}
\title{Poles and Poltergeists in $e^+ e^- \to D \bar D$ Data}
\author{Nils H\"{u}sken}
\email{nhuesken@uni-mainz.de}
\affiliation{Johannes Gutenberg University of Mainz, Johann-Joachim-Becher-Weg 45, D-55099 Mainz, Germany}
\author{Richard F. Lebed}
\email{Richard.Lebed@asu.edu}
\affiliation{Department of Physics, Arizona State University, Tempe,
AZ 85287, USA}
\author{Ryan~E.~Mitchell}
\email{remitche@iu.edu}
\affiliation{Indiana University, Bloomington, Indiana 47405, USA}
\author{Eric~S.~Swanson}
\email{swansone@pitt.edu}
\affiliation{University of Pittsburgh, Pittsburgh, PA 15260, USA}
\author{Ya-Qian~Wang}
\email{whyaqm@hbu.edu.cn}
\affiliation{Department of Physics, Hebei University, Baoding, 071002, China}
\author{Chang-Zheng Yuan}
\email{yuancz@ihep.ac.cn}
\affiliation{Institute of High Energy Physics, Chinese Academy of Sciences,
 Beijing 100049, China}
\affiliation{University of Chinese Academy of Sciences, Beijing 100049, China}

\date{April, 2024}

\begin{abstract}
A recent report of $e^+ e^- \to D\bar D$ events by the BESIII Collaboration suggests the presence of a structure $R$ at 3900~MeV\@.  We argue that this structure, called $G(3900)$ in the past, is not in fact due to a new $c\bar c$ resonance, but rather naturally emerges as a threshold enhancement due to the opening of the $D^*\bar D$ channel.  We further find that the appearance of this structure does not require suppression because of a radial node in the $\psi(4040)$ wave function, although a node improves fit quality.  The measured $e^+ e^-$ coupling of $\psi(4040)$ is found to be substantially smaller than previously estimated.  
In addition, we report new corrections to the measured cross section $\sigma(e^+ e^- \to D\bar D)$ at energies near $\psi(3770)$.

\end{abstract}

\maketitle

\section{Introduction}

The BESIII Collaboration recently presented results from a high-precision measurement of the exclusive process $e^+ e^- \to D \bar D$~\cite{BESIII:2024ths}.  Their report included a fit to the data with a coherent sum of eight Breit-Wigner amplitudes, one of which is a possible new resonance labeled as $R(3900)$.  Although the authors point out that the parameters of this purported resonance depend upon the chosen fit model, they nevertheless extract the values $m = 3872 \pm 14 \pm 3$ MeV and $\Gamma = 180 \pm 14 \pm 7$ MeV\@. 
This ostensible state has already been interpreted as a possible $D\bar D^*$ $P$-wave molecular resonance~\cite{Lin:2024qcq}.

The measured $e^+ e^- \to D \bar D$ cross section is compatible with earlier results from BaBar~\cite{BaBar:2006qlj,BaBar:2008drv} and Belle~\cite{Belle:2007qxm}.  Although these experiments see a clear peaking structure near $\sqrt{s} = 3900$~MeV, sometimes denoted $G(3900)$, the collaborations do not claim the presence of a new resonance, chiefly because this behavior was predicted more than forty years ago in a classic model calculation by the Cornell group~\cite{Eichten:1978tg}.  The model computes charmonium masses with non-relativistic kinematics and an assumed ``Coulomb-plus-linear" $c\bar c$ potential.  The authors then extend the model by coupling pure quarkonium states to continuum channels (in practice, $D^{(*)}\bar D^{(*)}$ and $D_s^{(*)}\bar D_s^{(*)}$), thereby permitting the computation of $e^+ e^- \to \mbox{hadrons}$ in the charm region~\cite{Eichten:1979ms}.  Their result in the $D\bar D$ channel shows a clear enhancement near 3900~GeV, which the authors attribute to the opening of the $D\bar D^* + D^*\bar D$ channel, as well as to the presence of a node in the $\psi(3S)$ wave function.

Here we examine this mechanism in a general framework to test its viability for explaining the recent \mbox{BESIII} data.
We find that a $K$-matrix fit can explain the enhancement, and that its primary cause is the opening of the $D^* \bar D$ channel, as predicted in Ref.~\cite{Eichten:1979ms}.  We also find that momentum dependence of the $\psi(4040)$ resonance coupling to the $D^*\bar D$ channel over hadronic scales is sufficient to drive the reaction rate down, thereby creating the enhancement that was fit as a Breit-Wigner resonance by the BESIII Collaboration.  
Our result implies that the appearance of a radial node in the $\psi(4040)$ wave function is not required to explain the steep dip observed in the cross section just above 4~GeV\@.  
Nevertheless, we find that fits with a modeled node do improve fit quality and conclude that, although not required to explain the data, there is weak evidence for the presence of a node in the decays of $\psi(4040)$.
Extracted pole positions are similar to those reported by the Particle Data Group (PDG)~\cite{ParticleDataGroup:2022pth}, but the width we obtain for $\psi(4040)$ is $\Gamma = 130\pm 30 \pm 125$~MeV (the second uncertainty reflecting variation over models), somewhat larger than the $80\pm 10$~MeV reported by the PDG\@.  We have also determined the $e^+ e^-$ coupling of $\psi(4040)$  to be $180 \pm 100 \pm 170$~eV, to be compared with the value $860 \pm 170$~eV reported by the PDG\@.

Previous work related to the questions addressed here includes a perturbative treatment of $\psi(2S)$-$\psi(3770)$ mixing in an effective Lagrangian approach~\cite{Zhang:2009gy}, wherein the authors treat $G(3900)$ as a resonance but find that it can also be explained by the $D^* \bar D$ threshold.  Du \textit{et al.}~\cite{Du:2016qcr} examine charm production in a similar mass range in  a Lippmann-Schwinger approach, using contact interactions that are constrained by heavy-quark symmetry, and find a pole near 3900 MeV\@.  Uglov \textit{et al.}~\cite{Uglov:2016orr} perform a $K$-matrix analysis of Belle data in the energy range 3.4--4.7~GeV\@.  This model describes the channels $D\bar{D}$, $D^* \bar{D}$, $D^*\bar{D^*}$, and $D\bar{D}\pi$ using 40 parameters, but no bare pole near 3900~MeV is found to be required.

Cao and Lenske~\cite{Cao:2014qna} examine the line shape of $\psi(3770)$ in a coupled-channel $T$-matrix approach,
finding good agreement with the data, although with a small pole mass of 3716(30)~MeV. They also conclude that the $G(3900)$ enhancement is due to a distortion of the $\psi(3770)$ tail by the $D\bar{D}^*$ threshold.

 The $K$-matrix formalism 
has recently been applied to the $\psi(3770)$ line shape~\cite{Hanhart:2023fud}.  This work concludes that BES(I,II,III) data are consistent with $\psi(3770)$ being due to a single pole, and that its non-$D\bar D$ branching fraction is less than 6\%, [the first version of this paper reported a larger branching fraction because the authors used  an  uncorrected $\sigma(e^+ e^- \to D\bar D)$ cross section].

Finally, von Detten~\textit{et al.}~\cite{vonDetten:2024eie} examine charmonium production near 4.2~GeV with a combination of perturbative and Dyson-equation methods, finding evidence for a resonance with a pole at $4227 - 25i$~MeV that is consistent with a $D_1 \bar D$ molecule.

In Sec.~\ref{sec:Model} we describe  the $K$-matrix fit model (introduced in Ref.~\cite{Husken:2022yik}) used here, and we present the results of the fits to $e^+ e^- \to D \bar D$ data in Sec.~\ref{sec:Results}. 
Section~\ref{sec:Concl} summarizes our findings and concludes.  Corrections to the measured cross sections for $\sigma(e^+ e^-\to D\bar D)$ are presented in the Appendix.

\section{$K$-matrix Model}
\label{sec:Model}

It is generally recognized that fitting complex data with an amplitude model that is a sum of Breit-Wigner resonances is inadequate in the case of overlapping resonances.  Indeed, an amplitude that is a naive sum of Breit-Wigner profiles does not respect unitarity.  This type of model is also inadequate when non-resonant behavior is important, such as when nearby thresholds can cause dramatic changes in event rates.  In view of charmed thresholds at $\bar D D^*$ (3871.7~MeV for $\bar D^0 D^{*0}$ and 3880.0~MeV for $D^- D^{*+}$) and $D_s^+ D_s^-$ (3936.7~MeV), it is prudent to construct an amplitude model that incorporates coupled channels and enforces unitarity.  We choose to employ the $K$-matrix model as implemented in Ref.~\cite{Husken:2022yik} for this purpose, presenting only the most crucial details here.

The $K$-matrix model couples resonances $R$ to two-body continuum channels $\mu$ via couplings $g_{R:\mu}$, and permits scattering between continuum channels via a contact-like term $f_{\mu:\nu}$, as follows:
\be
K_{\mu,\nu} = \sum_R \frac{g^{\vphantom\dagger}_{R:\mu} g^{\vphantom\dagger}_{R:\nu}}{m_R^2 -s} + f_{\mu,\nu}. \label{eq:ourK}
\ee
%
%
Guided by constituent quark models and phenomenology, the resonance couplings are written as 
\be
g^{\vphantom\dagger}_{R:\mu}(s) = \hat g^{\vphantom\dagger}_{R:\mu} \left(\frac{k_\mu(s)}{\beta}\right)^{\ell_\mu} \!\! \cdot \, \exp \left( - \frac{k^2_\mu(s)}{\beta^2} \right) .
\label{eq:gFormFactor}
\ee
%
%
Here, $\beta$ is a typical universal hadronic scale that is either fit to the data or fixed in different $K$-matrix models.  The relative momentum in the channel $\mu$ (with component masses $m_{1,2}$) is
\be
k_\mu(s) = \frac{1}{2\sqrt{s}} \sqrt{ \left[ s - (m_1 + m_2)^2 \right] \left[ s - (m_1 - m_2)^2 \right] } \, ,
\ee
and $\ell_\mu$ is the dominant orbital angular momentum in the channel $\mu$.

In view of the claims of the Cornell group, we have also fit the data with a coupling model that permits a radial node in the simplest possible way:
\be
g^{\textrm{node}}_{\psi(4040):\mu} = g_{\psi(4040):\mu}(s)\cdot \left( 1 - \frac{k_\mu^2(s)}{k_0^2} \right) ,
\label{eq:node}
\ee
where the node location $k_0$ is an additional fit parameter.

In a similar fashion, the non-resonant couplings $f$ are parameterized by 
\begin{align}
 f_{\mu ,\nu} = \hat{f}_{\mu , \nu}\cdot 
 \left(\frac{k_\mu(s)}{\beta}\right)^{\ell_\mu} \!\! \cdot 
 \left(\frac{k_\nu(s)}{\beta}\right)^{\ell_\nu} \nonumber  \\ \cdot \exp \left[- \, \frac{k^2_\mu(s)+ k^2_\nu(s)}{\beta^2} \right].
\label{eq:fFormFactor}
\end{align}
%
%
This form was chosen in Ref.~\cite{Husken:2022yik} for simplicity, consistency with the coupling model, and because the inclusion of  high-energy damping was found to be advantageous for fit robustness in the case of $B^{(*)}_{(s)} \bar B^{(*)}_{(s)}$ and $\Upsilon$ production.  The model was not applied to the $e^+ e^-$ channel, where a hadronic form factor is unnecessary.

In general, one expects the couplings $g^{\vphantom\dagger}_{R:\mu}$ to be isopin-independent, since the resonances themselves are isoscalar.  However, the continuum couplings $f_{\mu, \nu}$ for different charge states can vary substantially since they receive contributions from both the $I=0$ and $I = 1$ components of the photon, which can induce substantial isospin asymmetry through mixing with the threshold strong-interaction processes.  We will examine the effect of varying isospin constraints in the following discussion.

These model choices are inserted into the model-independent $K$-matrix formalism.  The scattering amplitude is written as
\be
\M = (1+KC)^{-1}K ,
\ee
%
%
where $C$ is the Chew-Mandelstam function that introduces the imaginary part of the scattering amplitude required for unitarity.  The specific form chosen for the Chew-Mandlestam function in this case~\cite{CrystalBarrel:2019zqh,Wilson:2014cna,Basdevant:1977ya} is designed to improve the analytic properties of the amplitude in the vicinity of thresholds.  The imaginary part of $C$ is given by
\be
\Im(C) = - \rho ,
\ee
%
%
where $\rho$ is a diagonal phase-space matrix in channel space, with elements given by
\be
\rho_{\mu,\nu} = \delta_{\mu,\nu} \frac{k_\nu(s)}{8 \pi  \sqrt{s}} .
\ee
%
%
Finally,  Aitchison's $P$-vector formalism~\cite{Aitchison:1972ay} is used to implement production via the $e^+ e^-$ channel:

\be
\M_{\mu,ee} = \sum_{\nu}(1+\hat K \hat C)^{-1}_{\mu,\nu} P_{\nu} ,
\ee
%
%
where  $P_\nu = K_{\nu,ee}$, while $\hat K$ and $\hat C$ are defined in the restricted channel space that excludes the initial channel.  Our fit models incorporate between 24 and 30 parameters, as reported in Table \ref{Tab:Models}.

\section{Fit Procedure and Results} 
\label{sec:Results}

Our choice for data sets to be fit, 
the specific implementation of the $K$-matrix model, and the fit procedure employed are described in the following subsections.

\subsection{Data Selection}
\label{sec:Data}

We choose to model the region $\sqrt{s} = 3.7$--$4.2$~GeV with two bare charmonium resonances, corresponding to $\psi(3770)$ and $\psi(4040)$ [nominally the $\psi(1D)$ and $\psi(3S)$ states of the Cornell model, respectively].  No additional or novel charmonia are assumed to contribute.  We truncate the data at 4200~MeV  to avoid complications  from known exotic states like $\psi(4230)$  and   from the effects of pairs of orbitally excited $D$ mesons such as $D_1(2420)$.  We have found it necessary to separate the $D^0 \bar D^0$ and $D^+ D^-$ data, presumably because the thresholds at 3730~MeV and 3739~MeV respectively, are sufficiently different to drive the differences visible in the cross-section data between the two channels, as discussed further below.

We also employ (and report in the Appendix) precise unpublished BESIII data for  $D\bar D$ cross sections in the vicinity of $\psi(3770)$~\cite{Julin:2017jcl}.
It is useful to remove initial-state radiation (ISR) and vacuum polarization (VP) effects from the data when fitting with the $K$-matrix model because the model is designed to extract hadronic information from the specific final state.  To this end, the observed cross section $\sigma^{\text{obs}}$ of $e^+ e^-\to D\bar D$ has been related to the dressed cross section $\sigma^{\text{dre}}$ by
\begin{equation}
   \sigma^{\text{dre}}=\frac{\sigma^{\text{obs}}}{1+\delta},
\end{equation}
where the ISR correction factor $1+\delta$ is obtained via an iterative procedure, following Ref.~\cite{DongXK}.
The Born cross section $\sigma^{\text{B}}$, which is the one used in our fits, is then calculated from the expression 
\begin{equation}    
   \sigma^{\text{B}}=\frac{\sigma^{\text{dre}}}{1/|1-\Pi|^2},
\end{equation}
where the VP factor $1/|1-\Pi|^2$ is taken from Ref.~\cite{rad2010}.

\begin{figure*}[htbp]  
\centering  
\includegraphics[width=0.45\textwidth]{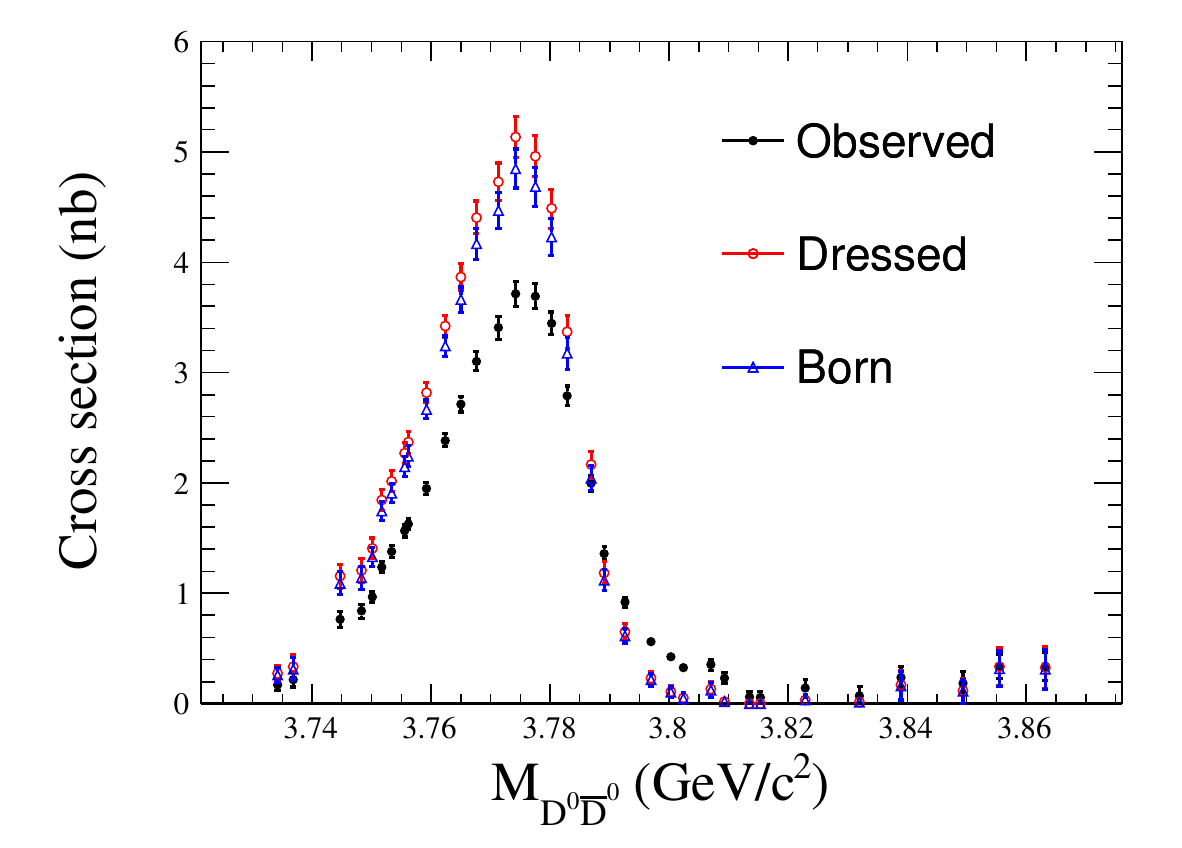}
\includegraphics[width=0.45\textwidth]{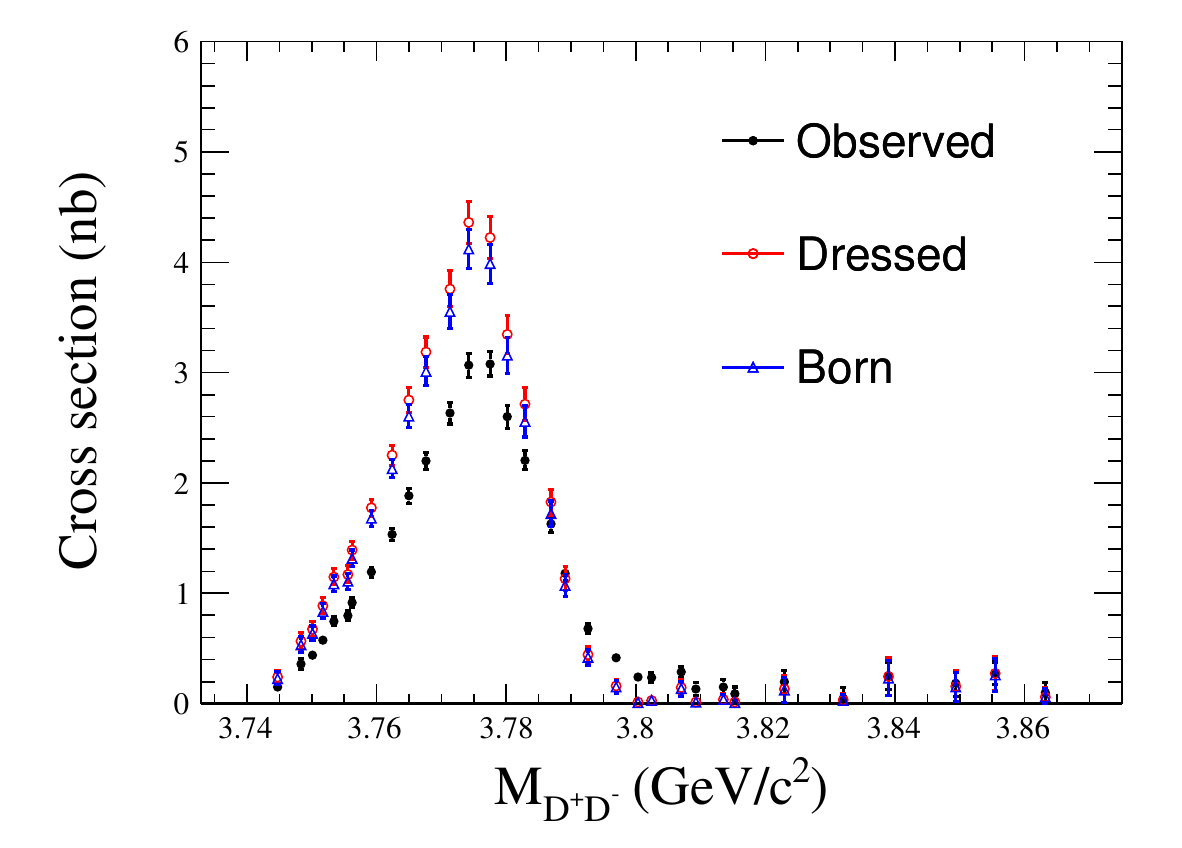}
\caption{Observed, dressed, and Born $e^+ e^-\to D\bar D$ cross sections in the vicinity of $\psi(3770)$.  Uncertainties are statistical only.}  
\label{fig:cs}  
\end{figure*}

Figure~\ref{fig:cs} displays the observed, dressed, and Born cross sections for the neutral and charged channels, respectively; uncertainties are statistical only.  A common systematic uncertainty of 3.22\% (3.17\%) is assigned to $e^+ e^-\to D^0\bar{D}^0$ ($e^+ e^-\to D^+D^-$)~\cite{Julin:2017jcl}.  The cross sections are compiled in Tables~\ref{tab:d0} and \ref{tab:dpm} in the Appendix.

In view of its claimed importance, the $D {\bar D}^* + D^* \bar{D}$ channel is also included in our model fit.  BESIII data~\cite{BESIII:2021yvc}
have been supplemented with data from Belle~\cite{Belle:2017grj} and CLEO-c~\cite{CLEO:2008ojp,Dong:2017tpt}, because BESIII measurements in this channel are sparsely spaced in energy.
 
Finally, if $D\bar D \! \leftrightarrow \! D\bar{D}^* \! + D^* \bar{D}$  rescattering does in fact turn out to be important, it is prudent to also consider the nearby $D^*\bar{D}^*$ channel in the coupled-channel fit model.  The data employed here are readily available from BESIII~\cite{BESIII:2021yvc}, 
 Belle~\cite{Belle:2017grj}, 
 and CLEO-c~\cite{CLEO:2008ojp,Dong:2017tpt}.

Previous work on bottom production has shown that incorporating the inclusive reaction $e^+ e^- \to b\bar b$ into the fit model is beneficial both because of the high quality of the data and the additional constraints it imposes upon the fit~\cite{Husken:2022yik}.  We have therefore incorporated an inclusive  $e^+ e^- \to c\bar c$ channel in this work.  The cross section was obtained by subtracting light-quark contributions to $R$ in the relevant mass region using Eqs.~(9.7)--(9.9) of the PDG review on Quantum Chromodynamics~\cite{ParticleDataGroup:2022pth}, with $\alpha_s(s)$ calculated using RunDec~\cite{Schmidt:2012az}, and multiplying by $\sigma(e^+ e^-\to \mu^+\mu^-)$.  While useful, a complication introduced by this method is that the sum of exclusive channels typically does not saturate the inclusive cross section.  This deficit is accounted for by introducing a ``dummy" channel---meant to represent all the neglected channels---that absorbs the difference in the rates, as discussed further in the next section.

We note that the $D_s^+ D_s^-$ channel opens at 3937~MeV, which provides a another potential source for a significant threshold effect.  However, we have chosen not to include this channel in the fit because of the paucity of available data. 

\subsection{Fit Procedure}

With these choices, the task is to fit a two-resonance, five-channel model to $D^0 \bar D^0$, $D^+ D^-$, $D \bar D^* + \bar D D^{*}$, $D^* \bar D^*$ (the latter two isospin-averaged), and the total cross-section data, along with the dummy channel.  In general, this task is a difficult one, and we have found it useful to implement the fit by progressively increasing the number of included channels.  Fits thus start with data exclusively from the $D\bar D$ channels, which forms the starting point for a three-channel fit to $D\bar D$ and $D \bar D^* + D^{*} \bar D$ data, and proceeds to the full five-channel model.

The need for a dummy channel to account for missing $c\bar c$ production has been discussed.  We have chosen to model this channel in two ways that appear most relevant: as a three-body channel $J/\psi \, \pi\pi$ treated as a two-body channel $J/\psi \, (\pi\pi)$ in a relative $S$ wave, or a $D_s^+ D_s^-$ channel in a relative $P$ wave.

The stability of the fit was tested by generating 200 pseudodata sets, fitting to these, and then using the resulting minimum of each fit as a new starting point for fits to the measured data.

In general, all model parameters are unconstrained; however, we have chosen to impose the resonance isospin constraint $g^{\vphantom\dagger}_{R:D^+D^-} = g^{\vphantom\dagger}_{R:D^0\bar{D^0}}$, which is expected to hold to good accuracy.  A similar constraint might be expected to hold for the cross-channel nonresonant couplings $f_{D^+D^-:\mu}$ and $f_{D^0\bar{D^0}:\mu}$, as well as for the elastic couplings $f_{D^+D^-:D^+D^-}$ and $f_{D^0\bar{D^0}:D^0\bar{D^0}}$.  In view of these possibilities, we have tested model sensitivity by fitting five models, varying the constraints and dummy channels as shown in Table \ref{Tab:Models}.

\begin{table}[h]
\caption{Model definitions.  Couplings $g$ and $f$ are defined in Eq.~(\ref{eq:ourK}), and the label ``dummy'' is defined in Sec.~\ref{sec:Data}.  $\# \, p$ indicates the number of parameters in each model.}
\label{Tab:Models}
    \begin{tabular*}{\columnwidth}{@{\extracolsep{\stretch{1}}}*{4}{c}@{}}
\hline\hline
Model \ & Channels & Constraints & \# $\! p$ \\
\hline
1 & $D\bar D$, $\bar D D^*$, $D^*\bar D^*$, & $g^{\vphantom\dagger}_{R:D^+D^-} = g^{\vphantom\dagger}_{R:D^0\bar D^0}$ & 29\\
  & \hspace{1em} $[J/\psi (\pi\pi)]_\textrm{dummy}$ & \\
2 & $D\bar D$, $\bar D D^*$, $D^*\bar D^*$, & $g^{\vphantom\dagger}_{R:D^+D^-} = g^{\vphantom\dagger}_{R:D^0\bar D^0}$ & 24\\
  & \hspace{1em} $[J/\psi (\pi\pi)]_\textrm{dummy}$
  & $f_{D^+D^-:\mu} = f_{D^0\bar D^0:\mu}$  \\
&  & $f_{D^+D^-:D^+D^-} = f_{D^0\bar D^0:D^0\bar D^0}$ & \\
3 & $D\bar D$, $\bar D D^*$, $D^*\bar D^*$, & $g^{\vphantom\dagger}_{R:D^+D^-} = g^{\vphantom\dagger}_{R:D^0\bar D^0}$ & 29 \\
& \hspace{1em} [$D_s^+ D_s^-]_\textrm{dummy}$ \\
4 & $D\bar D$, $\bar D D^*$, $D^*\bar D^*$, & $g^{\vphantom\dagger}_{R:D^+D^-} = g^{\vphantom\dagger}_{R:D^0\bar D^0}$ & 24 \\
 & \hspace{1em} [$D_s^+D_s^-]_\textrm{dummy}$ 
 &  $f_{D^+D^-:\mu} = f_{D^0\bar D^0:\mu}$ & \\
&  & $f_{D^+D^-:D^+D^-} = f_{D^0\bar D^0:D^0\bar D^0}$  & \\
 5 & \multicolumn{2}{c}{Model~1 plus node [Eq.~(\ref{eq:node})]} & 30 \\
\hline\hline
\end{tabular*}
\end{table}

Finally, the hadronic scale is set to $\beta= 1$~GeV, which was found to be optimal in terms of fit quality and stability in the case of bottom production~\cite{Husken:2022yik}.

\subsection{Fit Results}

\begin{figure*}[h]
\centering
\includegraphics[width=0.45\textwidth]{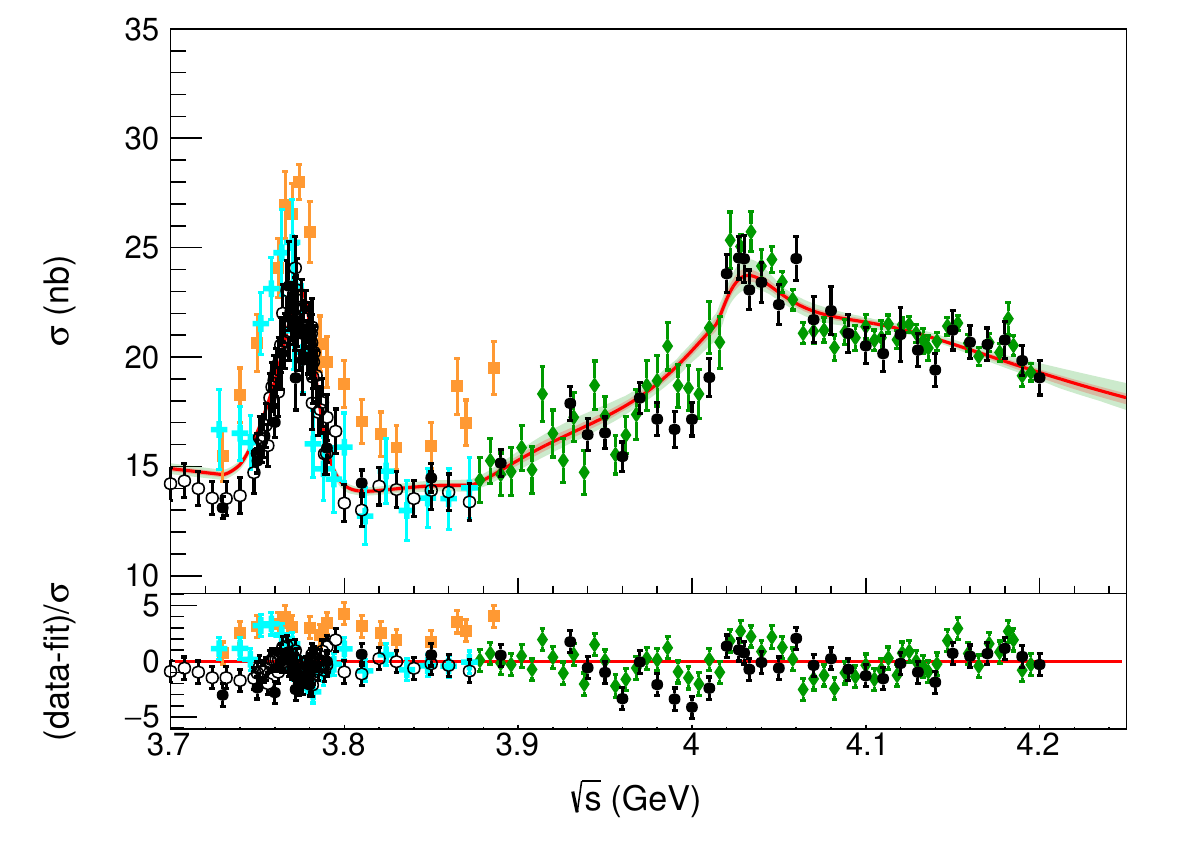} \qquad
    \includegraphics[width=.45\textwidth]{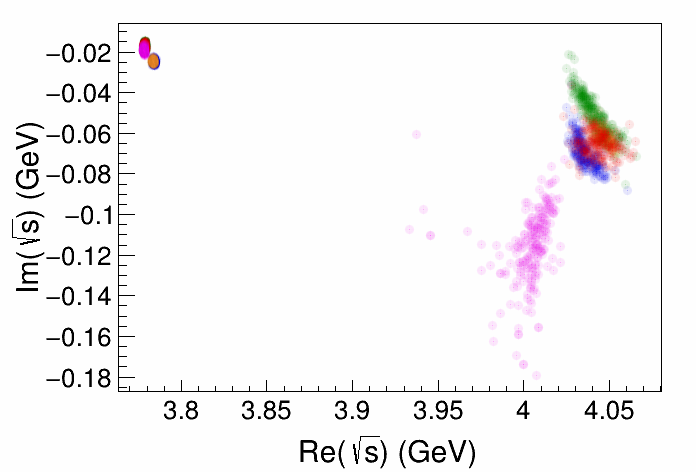}
\caption{Left: $e^+ e^- \to \textrm{hadrons}$ as fit using Model~1.  Data is from Refs.~\cite{BES:2001ckj} (BES, black closed), \cite{Ablikim:2006mb} (BESII, black open), \cite{Rapidis:1977cv} (SPEAR, orange), \cite{Osterheld:1986hw} (SPEAR, green), and \cite{Schindler:1979rj} (SPEAR, light blue).  Right:   Pole locations for Model~1 (red), Model~2 (blue), Model~3 (green), Model~4 (yellow), as well as Model~1 including a node (Model~5, pink).  Each point corresponds to a pole location in a fit to one of 200 sets of pseudodata.}
    \label{fig:3}   
\end{figure*}

All fit models reproduce the data fairly well in all channels, with the pull distributions indicating that the fairly large values of $\chi^{2}$/ndf are to a certain degree driven by mild inconsistencies between the experimental datasets (see, e.g., Fig. \ref{fig:3}, left).  In the two $D\bar D$ channels, where data solely from the BES\-III experiment is used, the fit quality is excellent.  Also note that we neglect mass differences between neutral and charged $D^{(*)}$ in $D^*\bar D + D \bar D^*$ and $D^*\bar D^*$, which can affect the fit quality for near-threshold data points.

\begin{figure*}[h]
\centering
\includegraphics[width=0.45\textwidth]{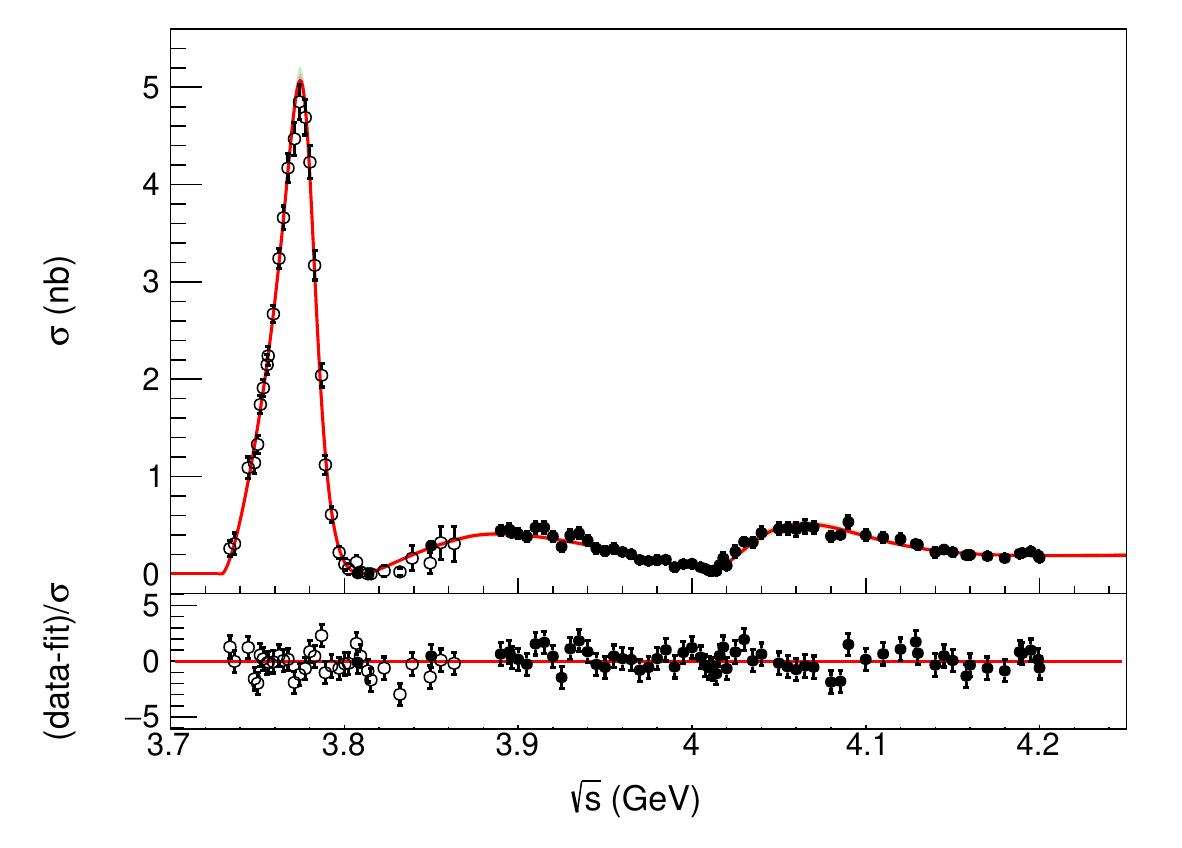} \qquad
    \includegraphics[width=.45\textwidth]{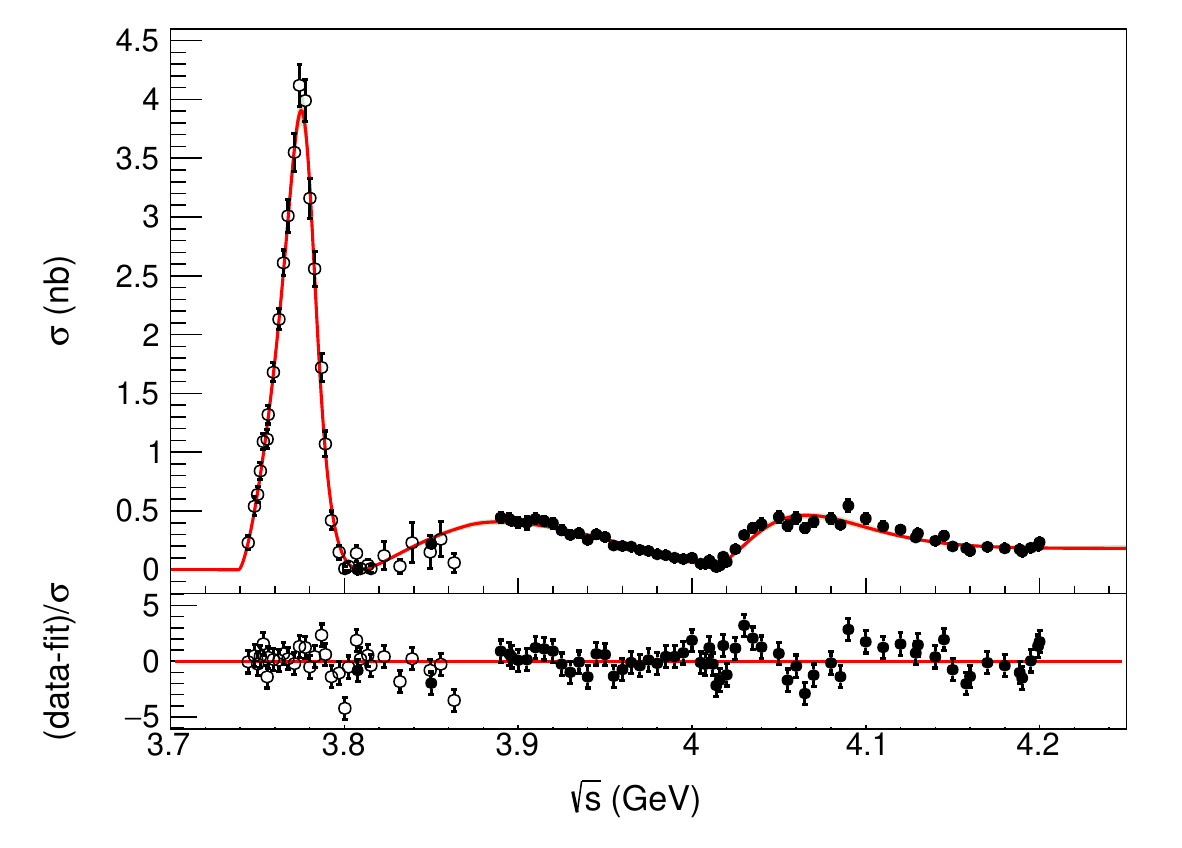}
\caption{Fit result for Model~1.  Left: $e^+ e^- \to D^0\bar D^0$.  Right:   $e^+ e^- \to D^+ D^-$.  Open data points are the Born cross section values based on observed cross sections, as reported in Ref.~\cite{Julin:2017jcl}; closed data points are from Ref.~\cite{BESIII:2024ths}.}
    \label{fig:1}   
\end{figure*}

\begin{figure*}[h]
\centering
\includegraphics[width=0.45\textwidth]{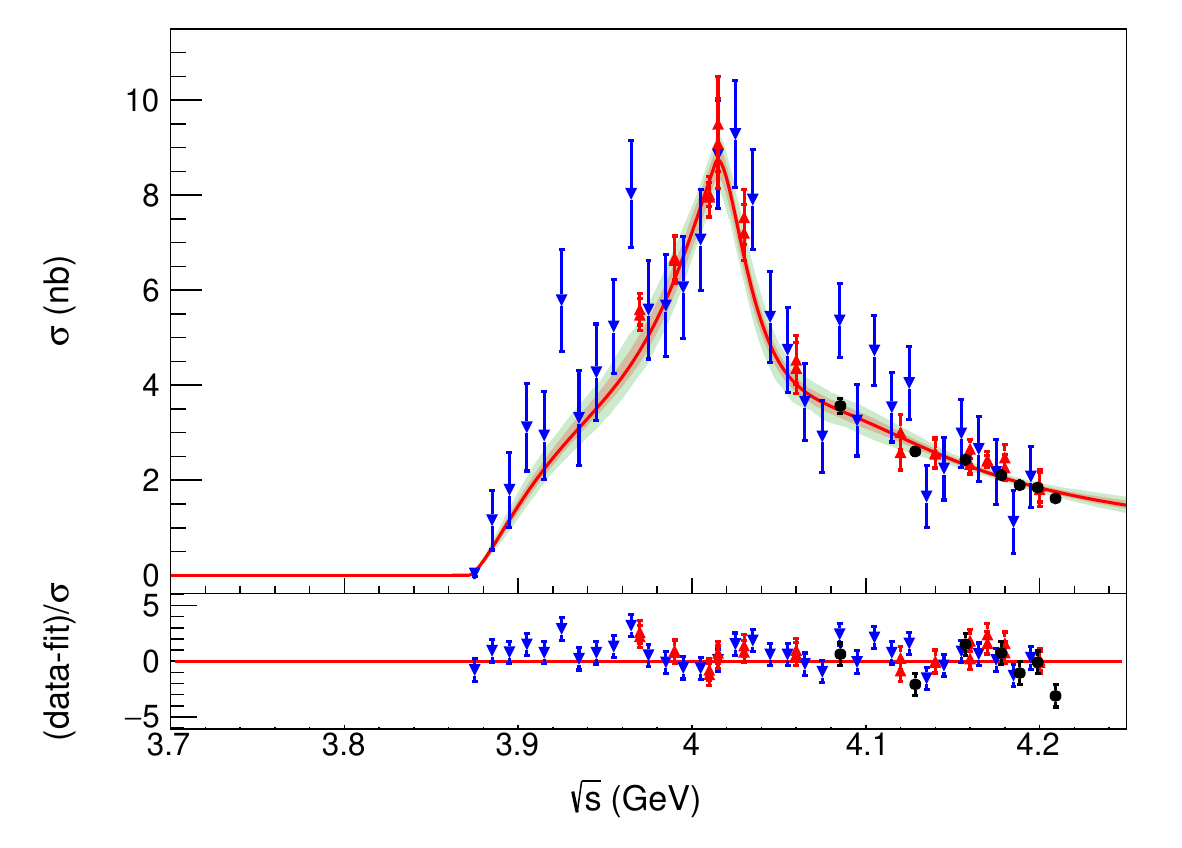} \qquad
    \includegraphics[width=.45\textwidth]{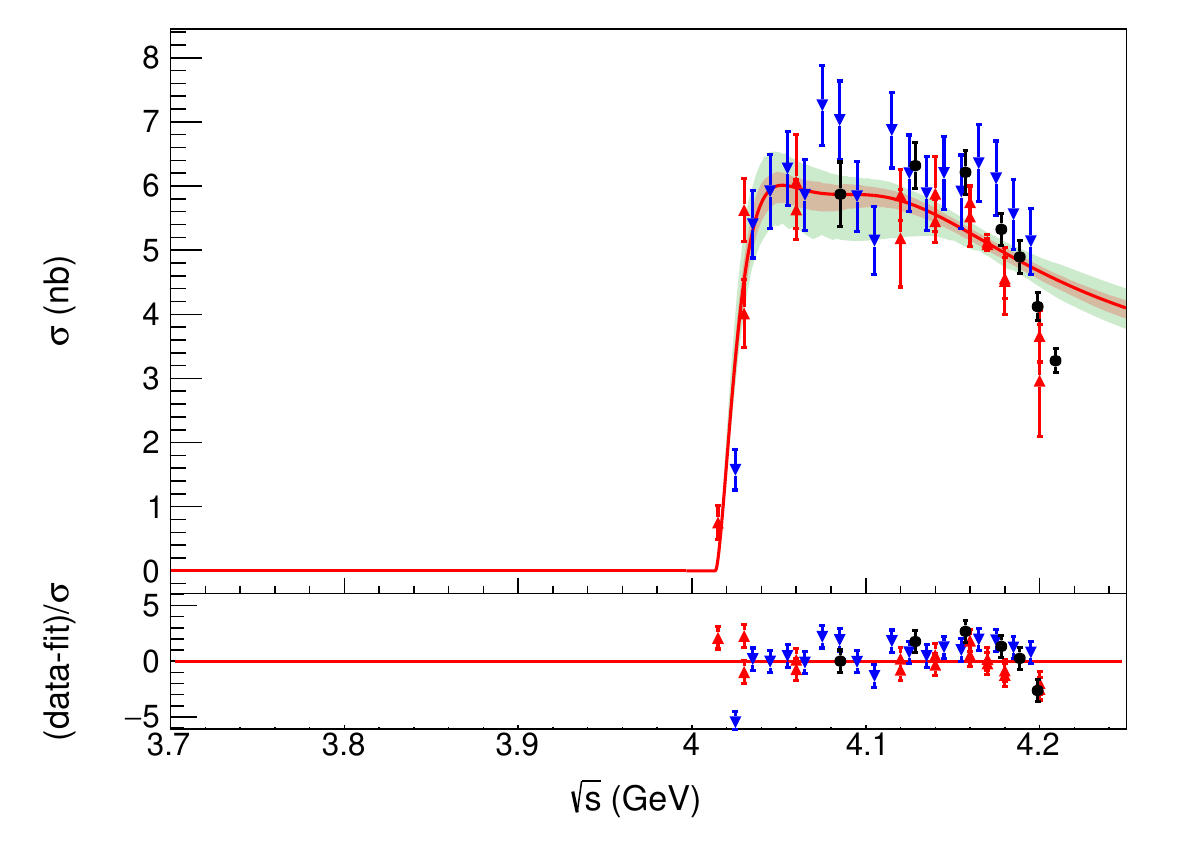}
\caption{Fit result for Model~1.  Left: $e^+ e^- \to D^*\bar D$.  Right: $e^+ e^- \to D^*\bar D^*$.  The red region indicates the 68\% confidence level, while green is the 90\% confidence level.  Black data points are from BESIII~\cite{BESIII:2021yvc}, red data is from CLEO-c~\cite{CLEO:2008ojp,Dong:2017tpt}, blue data is from Belle~\cite{Belle:2017grj}.}
    \label{fig:2}   
\end{figure*}

Results for all the models are very similar, with the exception that the fits for the $D^{*} \bar{D}$ channel (Fig.~\ref{fig:2}, left) that include a dummy $D_s^+ D_s^-$ channel (Models~3 and 4) exhibit a stronger rise near 3.95~GeV that better accommodates the high data points in this region.  As it gives the smallest overall $\chi^{2}$ for those models not containing a node, the fit results for Model~1 are shown in Figs.~\ref{fig:1}, \ref{fig:2}, and \ref{fig:3}.  The red area in Figs.~\ref{fig:2}--\ref{fig:3} represents 68\% confidence levels, while green indicates 90\% confidence levels.
We also note that, while the data in the range up to 4.2~GeV is described very well, the predictive power of the $K$-matrix model is very limited.  Extending the model to higher center-of-mass (c.m.) energies will certainly require additional channels and bare resonances.

We have extracted resonance parameters by analytically continuing the fit model and numerically searching for poles on the closest physical sheets (defined by negative imaginary part of the relative momentum $k_\mu(z)$ for channels below the real part of the pole mass, and positive for channels above). 
Partial widths are obtained by determining the residue at the pole in the relevant channel with the aid of Cauchy's theorem.  Our results are presented in Table~\ref{Tab:results} and Fig.~\ref{fig:3} (right).  The latter shows 200 pole locations for each of the five models, as determined by repeating the fits over 200 sets of pseudodata.

\begin{table}[h]
\caption{Extracted resonance parameters for Models~1--5 (defined in Table~\ref{Tab:Models}) and reported values by the PDG~\cite{ParticleDataGroup:2022pth}. \label{Tab:results}}
    \begin{tabular*}{\columnwidth}{@{\extracolsep{\stretch{1}}}*{6}{c}@{}}
\hline\hline
Model & Mass (MeV) & $\Gamma$ (MeV) & $\Gamma_{ee}$ (eV) & $\chi^2$/ndf\\
\hline
1 & 3778.7(7) & 34(4) & 205(25) & 2.20 \\
2 & 3784.2(7) & 49(4) & 3000(1500) & 2.48 \\
3 & 3778.9(6) & 33(4) & 210(20) & 2.39\\
4 & 3783.7(6) & 49(4) & 270(25) & 2.45\\
5 & 3778.3(6) & 38(5) & 200(400) & 1.88 \\
fit summary & 3778.7(7)(50) & 34(4)(15) & 205(25)(70) \\
PDG & 3773.7(4) & 27.2(1) & 261(21) \\
\hline
1 & 4044.0(15) & 130(30) & 180(100) \\
2 & 4036.0(10) & 135(35) & 15000(8000) \\
3 & 4040.0(10) & 95(30) & 80(80) \\
4 & 4046.0(10) & 120(20) & 10(50) \\
5 & 4008.0(10) & 220(80) & 50000(40000) \\
fit summary & 4044(15)(36) & 130(30)(125) & 180(100)(170) \\
PDG & 4039(1) & 80(10) & 856(162) \\
\hline\hline
\end{tabular*}
\end{table}

As noted above, the Cornell group suggested~\cite{Eichten:1979ms} that a node in the decay amplitudes of $\psi(4040)$ can affect open-charm production in the mass range considered here.  The good fit quality of Models 1--4 indicates that such a node is not required.  We nevertheless consider a fifth model that replicates Model 1 with modified resonance couplings, as shown in Eq.~(\ref{eq:node}).  We find that the resulting fit quality is somewhat improved, with $\chi^2$/ndf decreasing from 2.20 to 1.88.  Thus, we find weak evidence in favor of a node in the decay amplitudes of $\psi(4040)$.\footnote{A node in the amplitude for $\psi(4040) \to D\bar{D}$ implies similar nodes in $\psi(4040) \to D^{(*)}\bar{D}^{(*)}$ under very general assumptions~\cite{Barnes:1996ff}.}  We find it significant that our best-fit node parameter is $k_0 = 620$~MeV, remarkably close to the value of 750~MeV anticipated by the Cornell group~\cite{Eichten:1979ms}.

\section{Discussion and Conclusions} 
\label{sec:Concl}

We have performed a coupled-channel $K$-matrix fit
to the $D^0\bar D^0$, $D^+ D^- $, $D^* \bar{D}$ + c.c., $D^*\bar{D}^*$, and inclusive hidden-charm cross sections from $e^+ e^-$ for 3.7--4.2~GeV, and find a good fit in all channels for a variety of models.  The fit only assumes two bare poles, corresponding to $\psi(3770)$ and $\psi(4040)$.  The excellent quality of all the model fits  
is a strong indication that an additional bare pole near 3900 MeV is not needed to explain the data, in keeping with an old prediction from the Cornell group~\cite{Eichten:1979ms}.  This observation highlights the need for coupled-channel analyses when interpreting and fitting data in regions where thresholds are expected.

Although a node in the decay amplitude for $\psi(4040) \to D\bar D$ is expected on general grounds, we have found that it is not required to explain the enhancement at 3900~MeV\@, likely because  the $D^* \bar D$ threshold enhancement in combination with hadronic couplings that decrease over typical hadronic scales is sufficient to create the observed line shape.  Nevertheless, a model with $\psi(4040)$ couplings including a node does produce a fit with a somewhat better $\chi^2$/ndf, providing weak evidence for this mechanism and its effects on hadronic observables.

Resonance parameters are reported in Table \ref{Tab:results} for the different models.  The rows labeled ``fit summary" are provided for convenience, as a summary of this work.  It would be inappropriate to perform a statistical model averaging of our results, so we simply report our preferred result (Model 1) with an additional ``systematic uncertainty".  We find that the hadronic and electronic widths of $\psi(3770)$ are in most cases very close to those reported by the PDG\@.  It is interesting, however, that our preferred mass is 5~MeV above the PDG value---a significant difference, in view of the small uncertainties quoted.  Indeed, we have found that forcing the non-resonant couplings to be equal (Models 2 and 4) raises the $\psi(3770)$ pole location substantially. 

Similarly, the preferred $\psi(4040)$ mass and hadronic widths lie close to PDG values; however, the extracted electronic partial width (for Models~1, 3, and 4) is substantially smaller.
 This discrepancy is not due to a statistical fluctuation; rather, the PDG value is obtained from analyses that assume all of the structure in the total hadronic cross section to arise from resonances ({\it e.g.}, from a model consisting of a sum of Breit-Wigner amplitudes).  Such a choice, which ignores non-resonant contributions and rescattering, necessarily overestimates the electronic width.
In our model, substantial structure is generated by nonperturbative effects and by non-resonant scattering, naturally leading to a smaller residue and partial width.  Thus, we recommend that $\Gamma_{ee}[\psi(4040)]$ be updated to a smaller value, as obtained here.  We remark that similar problems affect bottomonium vectors, as reported in Ref.~\cite{Husken:2022yik}.

We note that Models 2 and 5 yield electronic widths that greatly deviate from the other models.  This discrepancy appears to be related to bootstrap fits in which the two bare pole positions that ultimately produce $\psi(3770)$ and $\psi(4040)$ unexpectedly move close to each other, or in which the scattering couplings become anomalously large (perhaps indicating that the models are sufficiently complex as to over-fit the data). 

In summary, a unitary coupled-channel analysis of charm production near 4.0~GeV finds strong evidence for the well-established $\psi(3770)$ and $\psi(4040)$ resonances, with no need for additional poles.  The extracted resonance parameters are in broad agreement with PDG values; however, we find a slightly higher $\psi(3770)$ pole mass at 3778.7(7)(50) MeV, and a lower electronic width for the $\psi(4040)$ of 180(100)(170) eV. Our results are in agreement with the predictions of the Cornell group, and affirm the importance of performing coupled-channel analyses when fitting data with overlapping resonances or nearby thresholds.

\begin{acknowledgments}
The authors acknowledge support from the European Union Horizon 2020 research and innovation programme under Marie Sk\l{}odowska-Curie grant agreement No.\ 894790, and the Helmholtz-Institut Mainz, Section SPECF (H\"usken), 
the National Science Foundation under Grant No.\ PHY-2110278 (Lebed), 
the U.S.\ Department of Energy under Contract No.\ DE-FG02-05ER41374 (Mitchell),
the U.S.\ Department of Energy under grant No.\ 
DE-SC0019232  (Swanson),
the National Key Research and Development Program of China under Contract No.~2020YFA0406300, and National Natural Science Foundation of China (NSFC) under contract No.~12361141819 (Yuan).
\end{acknowledgments}

\appendix
\section{$D\bar D$ Cross Sections}

Details and numerical results for the new BESIII measurement of the observed, dressed, and Born cross sections for $e^+ e^- \to D\bar D$ are presented here.
 
We remark that statistical uncertainties in the dressed and Born cross sections are not independent.  To account for this issue, 10,000 pseudo cross sections were generated at each energy point.  An iterative method (Ref.~\cite{Dong:2017tpt}) was then employed to extract the dressed and Born cross sections for each of the pseudodata sets.
 
 The resulting average cross section and the root of the diagonal elements of the covariance matrix are reported in the tables as the central value and the uncertainty for each entry.
In practice, we found it unnecessary to include the full covariance matrix in the fits, since off-diagonal correlations are negligible.

\begin{table}[ht]
    \centering
    \caption{Numerical results for $\sigma ( e^+ e^-\to D^0\bar{D}^0 )$ (in nb) as a function of c.m.\ energy $E$, including only statistical uncertainties.  The column entries are defined in Sec.~\ref{sec:Data}.}
    \begin{tabular*}{\columnwidth}{@{\extracolsep{\stretch{1}}}*{5}{c}@{}}
    \hline\hline
    $E$ (GeV)&Observed&Dressed&Born&VP\\
    \hline
3.7342 &  0.169 $\pm$  0.046 &  0.27 $\pm$  0.08 &  0.26 $\pm$  0.08 &  1.07 \\
3.7368 &  0.218 $\pm$  0.068 &  0.33 $\pm$  0.12 &  0.31 $\pm$  0.11 &  1.06 \\
3.7447 &  0.765 $\pm$  0.070 &  1.15 $\pm$  0.11 &  1.09 $\pm$  0.11 &  1.06 \\
3.7483 &  0.838 $\pm$  0.064 &  1.21 $\pm$  0.11 &  1.14 $\pm$  0.11 &  1.06 \\
3.7501 &  0.968 $\pm$  0.053 &  1.41 $\pm$  0.10 &  1.33 $\pm$  0.09 &  1.06 \\
3.7517 &  1.237 $\pm$  0.052 &  1.85 $\pm$  0.10 &  1.74 $\pm$  0.09 &  1.06 \\
3.7534 &  1.377 $\pm$  0.051 &  2.02 $\pm$  0.10 &  1.91 $\pm$  0.09 &  1.06 \\
3.7556 &  1.566 $\pm$  0.055 &  2.27 $\pm$  0.10 &  2.15 $\pm$  0.10 &  1.06 \\
3.7562 &  1.629 $\pm$  0.051 &  2.37 $\pm$  0.10 &  2.24 $\pm$  0.10 &  1.06 \\
3.7592 &  1.948 $\pm$  0.052 &  2.82 $\pm$  0.09 &  2.67 $\pm$  0.09 &  1.06 \\
3.7624 &  2.385 $\pm$  0.058 &  3.42 $\pm$  0.10 &  3.24 $\pm$  0.10 &  1.06 \\
3.7650 &  2.715 $\pm$  0.071 &  3.87 $\pm$  0.13 &  3.66 $\pm$  0.12 &  1.06 \\
3.7676 &  3.102 $\pm$  0.087 &  4.41 $\pm$  0.15 &  4.17 $\pm$  0.15 &  1.06 \\
3.7713 &  3.406 $\pm$  0.101 &  4.73 $\pm$  0.17 &  4.47 $\pm$  0.17 &  1.06 \\
3.7742 &  3.714 $\pm$  0.111 &  5.14 $\pm$  0.19 &  4.85 $\pm$  0.18 &  1.06 \\
3.7775 &  3.692 $\pm$  0.111 &  4.96 $\pm$  0.19 &  4.69 $\pm$  0.18 &  1.06 \\
3.7802 &  3.444 $\pm$  0.104 &  4.49 $\pm$  0.18 &  4.23 $\pm$  0.17 &  1.06 \\
3.7829 &  2.791 $\pm$  0.090 &  3.37 $\pm$  0.16 &  3.17 $\pm$  0.15 &  1.06 \\
3.7869 &  1.995 $\pm$  0.072 &  2.17 $\pm$  0.12 &  2.04 $\pm$  0.12 &  1.06 \\
3.7891 &  1.361 $\pm$  0.058 &  1.19 $\pm$  0.11 &  1.12 $\pm$  0.10 &  1.06 \\
3.7926 &  0.920 $\pm$  0.045 &  0.65 $\pm$  0.08 &  0.61 $\pm$  0.08 &  1.06 \\
3.7970 &  0.562 $\pm$  0.037 &  0.23 $\pm$  0.07 &  0.22 $\pm$  0.06 &  1.06 \\
3.8003 &  0.424 $\pm$  0.035 &  0.11 $\pm$  0.06 &  0.10 $\pm$  0.06 &  1.06 \\
3.8024 &  0.325 $\pm$  0.038 &  0.05 $\pm$  0.06 &  0.05 $\pm$  0.06 &  1.06 \\
3.8070 &  0.352 $\pm$  0.050 &  0.13 $\pm$  0.08 &  0.12 $\pm$  0.07 &  1.06 \\
3.8093 &  0.233 $\pm$  0.050 &  0.02 $\pm$  0.04 &  0.02 $\pm$  0.04 &  1.06 \\
3.8135 &  0.060 $\pm$  0.049 &  0.00 $\pm$  0.01 &  0.00 $\pm$  0.01 &  1.06 \\
3.8153 &  0.056 $\pm$  0.051 &  0.00 $\pm$  0.01 &  0.00 $\pm$  0.01 &  1.06 \\
3.8229 &  0.140 $\pm$  0.083 &  0.03 $\pm$  0.06 &  0.03 $\pm$  0.06 &  1.05 \\
3.8320 &  0.069 $\pm$  0.086 &  0.02 $\pm$  0.04 &  0.02 $\pm$  0.04 &  1.05 \\
3.8390 &  0.237 $\pm$  0.105 &  0.17 $\pm$  0.14 &  0.16 $\pm$  0.13 &  1.05 \\
3.8494 &  0.186 $\pm$  0.104 &  0.12 $\pm$  0.12 &  0.11 $\pm$  0.11 &  1.05 \\
3.8555 &  0.337 $\pm$  0.111 &  0.33 $\pm$  0.17 &  0.32 $\pm$  0.17 &  1.05 \\
3.8632 &  0.340 $\pm$  0.127 &  0.33 $\pm$  0.19 &  0.31 $\pm$  0.18 &  1.05 \\
    \hline\hline
    \end{tabular*}
        \label{tab:d0}
\end{table}
\clearpage
\begin{table}
    \centering
    \caption{Numerical results for $\sigma ( e^+ e^-\to D^+D^- )$ (in nb) as a function of c.m.\ energy $E$, including only statistical uncertainties.  The column entries are defined in Sec.~\ref{sec:Data}.}
    \begin{tabular*}{\columnwidth}{@{\extracolsep{\stretch{1}}}*{5}{c}@{}}
   \hline\hline
    $E$ (GeV)&Observed&Dressed&Born&VP\\
    \hline
3.7447 &  0.150 $\pm$  0.038 &  0.24 $\pm$  0.07 &  0.23 $\pm$  0.06 &  1.06 \\
3.7483 &  0.360 $\pm$  0.048 &  0.57 $\pm$  0.08 &  0.54 $\pm$  0.08 &  1.06 \\
3.7501 &  0.439 $\pm$  0.040 &  0.67 $\pm$  0.08 &  0.64 $\pm$  0.07 &  1.06 \\
3.7517 &  0.574 $\pm$  0.040 &  0.89 $\pm$  0.08 &  0.84 $\pm$  0.07 &  1.06 \\
3.7534 &  0.748 $\pm$  0.042 &  1.15 $\pm$  0.08 &  1.09 $\pm$  0.07 &  1.06 \\
3.7556 &  0.797 $\pm$  0.045 &  1.17 $\pm$  0.08 &  1.11 $\pm$  0.08 &  1.06 \\
3.7562 &  0.914 $\pm$  0.042 &  1.39 $\pm$  0.09 &  1.32 $\pm$  0.08 &  1.06 \\
3.7592 &  1.192 $\pm$  0.045 &  1.77 $\pm$  0.08 &  1.68 $\pm$  0.08 &  1.06 \\
3.7624 &  1.534 $\pm$  0.051 &  2.25 $\pm$  0.09 &  2.13 $\pm$  0.09 &  1.06 \\
3.7650 &  1.882 $\pm$  0.066 &  2.75 $\pm$  0.12 &  2.61 $\pm$  0.11 &  1.06 \\
3.7676 &  2.200 $\pm$  0.081 &  3.19 $\pm$  0.14 &  3.01 $\pm$  0.14 &  1.06 \\
3.7713 &  2.634 $\pm$  0.098 &  3.76 $\pm$  0.17 &  3.55 $\pm$  0.16 &  1.06 \\
3.7742 &  3.067 $\pm$  0.111 &  4.36 $\pm$  0.19 &  4.12 $\pm$  0.18 &  1.06 \\
3.7775 &  3.078 $\pm$  0.111 &  4.22 $\pm$  0.19 &  3.99 $\pm$  0.18 &  1.06 \\
3.7802 &  2.599 $\pm$  0.100 &  3.35 $\pm$  0.18 &  3.16 $\pm$  0.17 &  1.06 \\
3.7829 &  2.206 $\pm$  0.088 &  2.72 $\pm$  0.16 &  2.56 $\pm$  0.15 &  1.06 \\
3.7869 &  1.627 $\pm$  0.073 &  1.82 $\pm$  0.13 &  1.72 $\pm$  0.12 &  1.06 \\
3.7891 &  1.178 $\pm$  0.062 &  1.13 $\pm$  0.11 &  1.07 $\pm$  0.11 &  1.06 \\
3.7926 &  0.680 $\pm$  0.045 &  0.44 $\pm$  0.08 &  0.42 $\pm$  0.08 &  1.06 \\
3.7970 &  0.414 $\pm$  0.039 &  0.16 $\pm$  0.07 &  0.15 $\pm$  0.06 &  1.06 \\
3.8003 &  0.241 $\pm$  0.037 &  0.01 $\pm$  0.03 &  0.01 $\pm$  0.02 &  1.06 \\
3.8024 &  0.236 $\pm$  0.043 &  0.03 $\pm$  0.04 &  0.03 $\pm$  0.04 &  1.06 \\
3.8070 &  0.289 $\pm$  0.051 &  0.14 $\pm$  0.08 &  0.14 $\pm$  0.07 &  1.06 \\
3.8093 &  0.132 $\pm$  0.055 &  0.01 $\pm$  0.03 &  0.01 $\pm$  0.03 &  1.06 \\
3.8135 &  0.153 $\pm$  0.066 &  0.04 $\pm$  0.06 &  0.04 $\pm$  0.05 &  1.06 \\
3.8153 &  0.089 $\pm$  0.063 &  0.01 $\pm$  0.03 &  0.01 $\pm$  0.03 &  1.06 \\
3.8229 &  0.197 $\pm$  0.107 &  0.13 $\pm$  0.13 &  0.12 $\pm$  0.12 &  1.05 \\
3.8320 &  0.046 $\pm$  0.099 &  0.03 $\pm$  0.06 &  0.03 $\pm$  0.06 &  1.05 \\
3.8390 &  0.254 $\pm$  0.124 &  0.24 $\pm$  0.17 &  0.23 $\pm$  0.17 &  1.05 \\
3.8494 &  0.186 $\pm$  0.118 &  0.16 $\pm$  0.14 &  0.15 $\pm$  0.14 &  1.05 \\
3.8555 &  0.273 $\pm$  0.104 &  0.27 $\pm$  0.16 &  0.26 $\pm$  0.15 &  1.05 \\
3.8632 &  0.099 $\pm$  0.091 &  0.06 $\pm$  0.09 &  0.06 $\pm$  0.08 &  1.05 \\
    \hline\hline
    \end{tabular*}
        \label{tab:dpm}
\end{table}

\bibliographystyle{apsrev4-2}

\bibliography{charm}

\begin{thebibliography}{33}%
\makeatletter
\providecommand \@ifxundefined [1]{%
 \@ifx{#1\undefined}
}%
\providecommand \@ifnum [1]{%
 \ifnum #1\expandafter \@firstoftwo
 \else \expandafter \@secondoftwo
 \fi
}%
\providecommand \@ifx [1]{%
 \ifx #1\expandafter \@firstoftwo
 \else \expandafter \@secondoftwo
 \fi
}%
\providecommand \natexlab [1]{#1}%
\providecommand \enquote  [1]{``#1''}%
\providecommand \bibnamefont  [1]{#1}%
\providecommand \bibfnamefont [1]{#1}%
\providecommand \citenamefont [1]{#1}%
\providecommand \href@noop [0]{\@secondoftwo}%
\providecommand \href [0]{\begingroup \@sanitize@url \@href}%
\providecommand \@href[1]{\@@startlink{#1}\@@href}%
\providecommand \@@href[1]{\endgroup#1\@@endlink}%
\providecommand \@sanitize@url [0]{\catcode `\\12\catcode `\$12\catcode
  `\&12\catcode `\#12\catcode `\^12\catcode `\_12\catcode `\%12\relax}%
\providecommand \@@startlink[1]{}%
\providecommand \@@endlink[0]{}%
\providecommand \url  [0]{\begingroup\@sanitize@url \@url }%
\providecommand \@url [1]{\endgroup\@href {#1}{\urlprefix }}%
\providecommand \urlprefix  [0]{URL }%
\providecommand \Eprint [0]{\href }%
\providecommand \doibase [0]{https://doi.org/}%
\providecommand \selectlanguage [0]{\@gobble}%
\providecommand \bibinfo  [0]{\@secondoftwo}%
\providecommand \bibfield  [0]{\@secondoftwo}%
\providecommand \translation [1]{[#1]}%
\providecommand \BibitemOpen [0]{}%
\providecommand \bibitemStop [0]{}%
\providecommand \bibitemNoStop [0]{.\EOS\space}%
\providecommand \EOS [0]{\spacefactor3000\relax}%
\providecommand \BibitemShut  [1]{\csname bibitem#1\endcsname}%
\let\auto@bib@innerbib\@empty
\bibitem [{\citenamefont {Ablikim}\ \emph {et~al.}()\citenamefont {Ablikim}
  \emph {et~al.}}]{BESIII:2024ths}%
  \BibitemOpen
  \bibfield  {author} {\bibinfo {author} {\bibfnamefont {M.}~\bibnamefont
  {Ablikim}} \emph {et~al.} (\bibinfo {collaboration} {BESIII Collaboration}),\
  }\href@noop {} {\ }\Eprint {https://arxiv.org/abs/2402.03829}
  {arXiv:2402.03829 [hep-ex]} \BibitemShut {NoStop}%
\bibitem [{\citenamefont {Lin}\ \emph {et~al.}(2024)\citenamefont {Lin},
  \citenamefont {Wang}, \citenamefont {Cheng}, \citenamefont {Meng},\ and\
  \citenamefont {Zhu}}]{Lin:2024qcq}%
  \BibitemOpen
  \bibfield  {author} {\bibinfo {author} {\bibfnamefont {Z.-Y.}\ \bibnamefont
  {Lin}}, \bibinfo {author} {\bibfnamefont {J.-Z.}\ \bibnamefont {Wang}},
  \bibinfo {author} {\bibfnamefont {J.-B.}\ \bibnamefont {Cheng}}, \bibinfo
  {author} {\bibfnamefont {L.}~\bibnamefont {Meng}},\ and\ \bibinfo {author}
  {\bibfnamefont {S.-L.}\ \bibnamefont {Zhu}},\ }\href@noop {} {\  (\bibinfo
  {year} {2024})},\ \Eprint {https://arxiv.org/abs/2403.01727}
  {arXiv:2403.01727 [hep-ph]} \BibitemShut {NoStop}%
\bibitem [{\citenamefont {Aubert}\ \emph {et~al.}(2007)\citenamefont {Aubert}
  \emph {et~al.}}]{BaBar:2006qlj}%
  \BibitemOpen
  \bibfield  {author} {\bibinfo {author} {\bibfnamefont {B.}~\bibnamefont
  {Aubert}} \emph {et~al.} (\bibinfo {collaboration} {BaBar Collaboration}),\
  }\href {https://doi.org/10.1103/PhysRevD.76.111105} {\bibfield  {journal}
  {\bibinfo  {journal} {Phys.\ Rev.\ D}\ }\textbf {\bibinfo {volume} {{\bf
  76}}},\ \bibinfo {pages} {111105} (\bibinfo {year} {2007})},\ \Eprint
  {https://arxiv.org/abs/hep-ex/0607083} {arXiv:hep-ex/0607083} \BibitemShut
  {NoStop}%
\bibitem [{\citenamefont {Aubert}\ \emph {et~al.}(2008)\citenamefont {Aubert}
  \emph {et~al.}}]{BaBar:2008drv}%
  \BibitemOpen
  \bibfield  {author} {\bibinfo {author} {\bibfnamefont {B.}~\bibnamefont
  {Aubert}} \emph {et~al.} (\bibinfo {collaboration} {BaBar Collaboration}),\
  }\href@noop {} {\  (\bibinfo {year} {2008})},\ \Eprint
  {https://arxiv.org/abs/0710.1371} {arXiv:0710.1371 [hep-ex]} \BibitemShut
  {NoStop}%
\bibitem [{\citenamefont {Pakhlova}\ \emph {et~al.}(2008)\citenamefont
  {Pakhlova} \emph {et~al.}}]{Belle:2007qxm}%
  \BibitemOpen
  \bibfield  {author} {\bibinfo {author} {\bibfnamefont {G.}~\bibnamefont
  {Pakhlova}} \emph {et~al.} (\bibinfo {collaboration} {Belle Collaboration}),\
  }\href {https://doi.org/10.1103/PhysRevD.77.011103} {\bibfield  {journal}
  {\bibinfo  {journal} {Phys.\ Rev.\ D}\ }\textbf {\bibinfo {volume} {{\bf
  77}}},\ \bibinfo {pages} {011103} (\bibinfo {year} {2008})},\ \Eprint
  {https://arxiv.org/abs/0708.0082} {arXiv:0708.0082 [hep-ex]} \BibitemShut
  {NoStop}%
\bibitem [{\citenamefont {Eichten}\ \emph {et~al.}(1978)\citenamefont
  {Eichten}, \citenamefont {Gottfried}, \citenamefont {Kinoshita},
  \citenamefont {Lane},\ and\ \citenamefont {Yan}}]{Eichten:1978tg}%
  \BibitemOpen
  \bibfield  {author} {\bibinfo {author} {\bibfnamefont {E.}~\bibnamefont
  {Eichten}}, \bibinfo {author} {\bibfnamefont {K.}~\bibnamefont {Gottfried}},
  \bibinfo {author} {\bibfnamefont {T.}~\bibnamefont {Kinoshita}}, \bibinfo
  {author} {\bibfnamefont {K.}~\bibnamefont {Lane}},\ and\ \bibinfo {author}
  {\bibfnamefont {T.-M.}\ \bibnamefont {Yan}},\ }\href
  {https://doi.org/10.1103/PhysRevD.17.3090} {\bibfield  {journal} {\bibinfo
  {journal} {Phys.\ Rev.\ D}\ }\textbf {\bibinfo {volume} {{\bf 17}}},\
  \bibinfo {pages} {3090} (\bibinfo {year} {1978})},\ \bibinfo {note}
  {[Erratum: Phys.\ Rev.\ D {\bf 21}, 313 (1980)]}\BibitemShut {NoStop}%
\bibitem [{\citenamefont {Eichten}\ \emph {et~al.}(1980)\citenamefont
  {Eichten}, \citenamefont {Gottfried}, \citenamefont {Kinoshita},
  \citenamefont {Lane},\ and\ \citenamefont {Yan}}]{Eichten:1979ms}%
  \BibitemOpen
  \bibfield  {author} {\bibinfo {author} {\bibfnamefont {E.}~\bibnamefont
  {Eichten}}, \bibinfo {author} {\bibfnamefont {K.}~\bibnamefont {Gottfried}},
  \bibinfo {author} {\bibfnamefont {T.}~\bibnamefont {Kinoshita}}, \bibinfo
  {author} {\bibfnamefont {K.}~\bibnamefont {Lane}},\ and\ \bibinfo {author}
  {\bibfnamefont {T.-M.}\ \bibnamefont {Yan}},\ }\href
  {https://doi.org/10.1103/PhysRevD.21.203} {\bibfield  {journal} {\bibinfo
  {journal} {Phys.\ Rev.\ D}\ }\textbf {\bibinfo {volume} {{\bf 21}}},\
  \bibinfo {pages} {203} (\bibinfo {year} {1980})}\BibitemShut {NoStop}%
\bibitem [{\citenamefont {Workman}\ \emph {et~al.}(2022)\citenamefont {Workman}
  \emph {et~al.}}]{ParticleDataGroup:2022pth}%
  \BibitemOpen
  \bibfield  {author} {\bibinfo {author} {\bibfnamefont {R.}~\bibnamefont
  {Workman}} \emph {et~al.} (\bibinfo {collaboration} {Particle Data Group}),\
  }\href {https://doi.org/10.1093/ptep/ptac097} {\bibfield  {journal} {\bibinfo
   {journal} {PTEP}\ }\textbf {\bibinfo {volume} {{\bf 2022}}},\ \bibinfo
  {pages} {083C01} (\bibinfo {year} {2022})}\BibitemShut {NoStop}%
\bibitem [{\citenamefont {Zhang}\ and\ \citenamefont
  {Zhao}(2010)}]{Zhang:2009gy}%
  \BibitemOpen
  \bibfield  {author} {\bibinfo {author} {\bibfnamefont {Y.-J.}\ \bibnamefont
  {Zhang}}\ and\ \bibinfo {author} {\bibfnamefont {Q.}~\bibnamefont {Zhao}},\
  }\href {https://doi.org/10.1103/PhysRevD.81.034011} {\bibfield  {journal}
  {\bibinfo  {journal} {Phys.\ Rev.\ D}\ }\textbf {\bibinfo {volume} {{\bf
  81}}},\ \bibinfo {pages} {034011} (\bibinfo {year} {2010})},\ \Eprint
  {https://arxiv.org/abs/0911.5651} {arXiv:0911.5651 [hep-ph]} \BibitemShut
  {NoStop}%
\bibitem [{\citenamefont {Du}\ \emph {et~al.}(2016)\citenamefont {Du},
  \citenamefont {Mei\ss{}ner},\ and\ \citenamefont {Wang}}]{Du:2016qcr}%
  \BibitemOpen
  \bibfield  {author} {\bibinfo {author} {\bibfnamefont {M.-L.}\ \bibnamefont
  {Du}}, \bibinfo {author} {\bibfnamefont {U.-G.}\ \bibnamefont
  {Mei\ss{}ner}},\ and\ \bibinfo {author} {\bibfnamefont {Q.}~\bibnamefont
  {Wang}},\ }\href {https://doi.org/10.1103/PhysRevD.94.096006} {\bibfield
  {journal} {\bibinfo  {journal} {Phys.\ Rev.\ D}\ }\textbf {\bibinfo {volume}
  {{\bf 94}}},\ \bibinfo {pages} {096006} (\bibinfo {year} {2016})},\ \Eprint
  {https://arxiv.org/abs/1608.02537} {arXiv:1608.02537 [hep-ph]} \BibitemShut
  {NoStop}%
\bibitem [{\citenamefont {Uglov}\ \emph {et~al.}(2017)\citenamefont {Uglov},
  \citenamefont {Kalashnikova}, \citenamefont {Nefediev}, \citenamefont
  {Pakhlova},\ and\ \citenamefont {Pakhlov}}]{Uglov:2016orr}%
  \BibitemOpen
  \bibfield  {author} {\bibinfo {author} {\bibfnamefont {T.}~\bibnamefont
  {Uglov}}, \bibinfo {author} {\bibfnamefont {Y.}~\bibnamefont {Kalashnikova}},
  \bibinfo {author} {\bibfnamefont {A.}~\bibnamefont {Nefediev}}, \bibinfo
  {author} {\bibfnamefont {G.}~\bibnamefont {Pakhlova}},\ and\ \bibinfo
  {author} {\bibfnamefont {P.}~\bibnamefont {Pakhlov}},\ }\href
  {https://doi.org/10.1134/S0021364017010064} {\bibfield  {journal} {\bibinfo
  {journal} {JETP Lett.}\ }\textbf {\bibinfo {volume} {{\bf 105}}},\ \bibinfo
  {pages} {1} (\bibinfo {year} {2017})},\ \Eprint
  {https://arxiv.org/abs/1611.07582} {arXiv:1611.07582 [hep-ph]} \BibitemShut
  {NoStop}%
\bibitem [{\citenamefont {Cao}\ and\ \citenamefont
  {Lenske}(2014)}]{Cao:2014qna}%
  \BibitemOpen
  \bibfield  {author} {\bibinfo {author} {\bibfnamefont {X.}~\bibnamefont
  {Cao}}\ and\ \bibinfo {author} {\bibfnamefont {H.}~\bibnamefont {Lenske}},\
  }\href@noop {} {\  (\bibinfo {year} {2014})},\ \Eprint
  {https://arxiv.org/abs/1410.1375} {arXiv:1410.1375 [nucl-th]} \BibitemShut
  {NoStop}%
\bibitem [{\citenamefont {Hanhart}\ \emph {et~al.}(2023)\citenamefont
  {Hanhart}, \citenamefont {K\"urten}, \citenamefont {Reboud},\ and\
  \citenamefont {van Dyk}}]{Hanhart:2023fud}%
  \BibitemOpen
  \bibfield  {author} {\bibinfo {author} {\bibfnamefont {C.}~\bibnamefont
  {Hanhart}}, \bibinfo {author} {\bibfnamefont {S.}~\bibnamefont {K\"urten}},
  \bibinfo {author} {\bibfnamefont {M.}~\bibnamefont {Reboud}},\ and\ \bibinfo
  {author} {\bibfnamefont {D.}~\bibnamefont {van Dyk}},\ }\href@noop {} {\
  (\bibinfo {year} {2023})},\ \Eprint {https://arxiv.org/abs/2312.00619}
  {arXiv:2312.00619 [hep-ph]} \BibitemShut {NoStop}%
\bibitem [{\citenamefont {von Detten}\ \emph {et~al.}(2024)\citenamefont {von
  Detten}, \citenamefont {Baru}, \citenamefont {Hanhart}, \citenamefont {Wang},
  \citenamefont {Winney},\ and\ \citenamefont {Zhao}}]{vonDetten:2024eie}%
  \BibitemOpen
  \bibfield  {author} {\bibinfo {author} {\bibfnamefont {L.}~\bibnamefont {von
  Detten}}, \bibinfo {author} {\bibfnamefont {V.}~\bibnamefont {Baru}},
  \bibinfo {author} {\bibfnamefont {C.}~\bibnamefont {Hanhart}}, \bibinfo
  {author} {\bibfnamefont {Q.}~\bibnamefont {Wang}}, \bibinfo {author}
  {\bibfnamefont {D.}~\bibnamefont {Winney}},\ and\ \bibinfo {author}
  {\bibfnamefont {Q.}~\bibnamefont {Zhao}},\ }\href@noop {} {\  (\bibinfo
  {year} {2024})},\ \Eprint {https://arxiv.org/abs/2402.03057}
  {arXiv:2402.03057 [hep-ph]} \BibitemShut {NoStop}%
\bibitem [{\citenamefont {H\"usken}\ \emph {et~al.}(2022)\citenamefont
  {H\"usken}, \citenamefont {Mitchell},\ and\ \citenamefont
  {Swanson}}]{Husken:2022yik}%
  \BibitemOpen
  \bibfield  {author} {\bibinfo {author} {\bibfnamefont {N.}~\bibnamefont
  {H\"usken}}, \bibinfo {author} {\bibfnamefont {R.}~\bibnamefont {Mitchell}},\
  and\ \bibinfo {author} {\bibfnamefont {E.}~\bibnamefont {Swanson}},\ }\href
  {https://doi.org/10.1103/PhysRevD.106.094013} {\bibfield  {journal} {\bibinfo
   {journal} {Phys.\ Rev.\ D}\ }\textbf {\bibinfo {volume} {{\bf 106}}},\
  \bibinfo {pages} {094013} (\bibinfo {year} {2022})},\ \Eprint
  {https://arxiv.org/abs/2204.11915} {arXiv:2204.11915 [hep-ph]} \BibitemShut
  {NoStop}%
\bibitem [{\citenamefont {Albrecht}\ \emph {et~al.}(2020)\citenamefont
  {Albrecht} \emph {et~al.}}]{CrystalBarrel:2019zqh}%
  \BibitemOpen
  \bibfield  {author} {\bibinfo {author} {\bibfnamefont {M.}~\bibnamefont
  {Albrecht}} \emph {et~al.} (\bibinfo {collaboration} {Crystal Barrel
  Collaboration}),\ }\href {https://doi.org/10.1140/epjc/s10052-020-7930-x}
  {\bibfield  {journal} {\bibinfo  {journal} {Eur.\ Phys.\ J. C}\ }\textbf
  {\bibinfo {volume} {{\bf 80}}},\ \bibinfo {pages} {453} (\bibinfo {year}
  {2020})},\ \Eprint {https://arxiv.org/abs/1909.07091} {arXiv:1909.07091
  [hep-ex]} \BibitemShut {NoStop}%
\bibitem [{\citenamefont {Wilson}\ \emph {et~al.}(2015)\citenamefont {Wilson},
  \citenamefont {Dudek}, \citenamefont {Edwards},\ and\ \citenamefont
  {Thomas}}]{Wilson:2014cna}%
  \BibitemOpen
  \bibfield  {author} {\bibinfo {author} {\bibfnamefont {D.}~\bibnamefont
  {Wilson}}, \bibinfo {author} {\bibfnamefont {J.}~\bibnamefont {Dudek}},
  \bibinfo {author} {\bibfnamefont {R.}~\bibnamefont {Edwards}},\ and\ \bibinfo
  {author} {\bibfnamefont {C.}~\bibnamefont {Thomas}},\ }\href
  {https://doi.org/10.1103/PhysRevD.91.054008} {\bibfield  {journal} {\bibinfo
  {journal} {Phys.\ Rev.\ D}\ }\textbf {\bibinfo {volume} {{\bf 91}}},\
  \bibinfo {pages} {054008} (\bibinfo {year} {2015})},\ \Eprint
  {https://arxiv.org/abs/1411.2004} {arXiv:1411.2004 [hep-ph]} \BibitemShut
  {NoStop}%
\bibitem [{\citenamefont {Basdevant}\ and\ \citenamefont
  {Berger}(1977)}]{Basdevant:1977ya}%
  \BibitemOpen
  \bibfield  {author} {\bibinfo {author} {\bibfnamefont {J.}~\bibnamefont
  {Basdevant}}\ and\ \bibinfo {author} {\bibfnamefont {E.}~\bibnamefont
  {Berger}},\ }\href {https://doi.org/10.1103/PhysRevD.16.657} {\bibfield
  {journal} {\bibinfo  {journal} {Phys.\ Rev.\ D}\ }\textbf {\bibinfo {volume}
  {{\bf 16}}},\ \bibinfo {pages} {657} (\bibinfo {year} {1977})}\BibitemShut
  {NoStop}%
\bibitem [{\citenamefont {Aitchison}(1972)}]{Aitchison:1972ay}%
  \BibitemOpen
  \bibfield  {author} {\bibinfo {author} {\bibfnamefont {I.}~\bibnamefont
  {Aitchison}},\ }\href {https://doi.org/10.1016/0375-9474(72)90305-3}
  {\bibfield  {journal} {\bibinfo  {journal} {Nucl.\ Phys.\ A}\ }\textbf
  {\bibinfo {volume} {{\bf 189}}},\ \bibinfo {pages} {417} (\bibinfo {year}
  {1972})}\BibitemShut {NoStop}%
\bibitem [{\citenamefont {Julin}(2017)}]{Julin:2017jcl}%
  \BibitemOpen
  \bibfield  {author} {\bibinfo {author} {\bibfnamefont {A.}~\bibnamefont
  {Julin}},\ }\emph {\bibinfo {title} {{\it Measurement of $D\bar{D}$ Decays
  from the $\psi(3770)$ Resonance}}},\ \href@noop {} {Ph.D. thesis},\ \bibinfo
  {school} {Minnesota U.} (\bibinfo {year} {2017})\BibitemShut {NoStop}%
\bibitem [{\citenamefont {Dong}\ \emph {et~al.}(2020)\citenamefont {Dong},
  \citenamefont {Mo}, \citenamefont {Wang},\ and\ \citenamefont
  {Yuan}}]{DongXK}%
  \BibitemOpen
  \bibfield  {author} {\bibinfo {author} {\bibfnamefont {X.-K.}\ \bibnamefont
  {Dong}}, \bibinfo {author} {\bibfnamefont {X.-H.}\ \bibnamefont {Mo}},
  \bibinfo {author} {\bibfnamefont {P.}~\bibnamefont {Wang}},\ and\ \bibinfo
  {author} {\bibfnamefont {C.-Z.}\ \bibnamefont {Yuan}},\ }\href
  {https://doi.org/10.1088/1674-1137/44/8/083001} {\bibfield  {journal}
  {\bibinfo  {journal} {Chin.\ Phys.\ C}\ }\textbf {\bibinfo {volume} {{\bf
  44}}},\ \bibinfo {pages} {083001} (\bibinfo {year} {2020})},\ \Eprint
  {https://arxiv.org/abs/2002.09838} {arXiv:2002.09838 [hep-ph]} \BibitemShut
  {NoStop}%
\bibitem [{\citenamefont {Actis}\ \emph {et~al.}(2010)\citenamefont {Actis}
  \emph {et~al.}}]{rad2010}%
  \BibitemOpen
  \bibfield  {author} {\bibinfo {author} {\bibfnamefont {S.}~\bibnamefont
  {Actis}} \emph {et~al.} (\bibinfo {collaboration} {Working Group on Radiative
  Corrections, Monte Carlo Generators for Low Energies}),\ }\href
  {https://doi.org/10.1140/epjc/s10052-010-1251-4} {\bibfield  {journal}
  {\bibinfo  {journal} {Eur. Phys. J. C}\ }\textbf {\bibinfo {volume} {66}},\
  \bibinfo {pages} {585} (\bibinfo {year} {2010})},\ \Eprint
  {https://arxiv.org/abs/0912.0749} {arXiv:0912.0749 [hep-ph]} \BibitemShut
  {NoStop}%
\bibitem [{\citenamefont {Ablikim}\ \emph {et~al.}(2022)\citenamefont {Ablikim}
  \emph {et~al.}}]{BESIII:2021yvc}%
  \BibitemOpen
  \bibfield  {author} {\bibinfo {author} {\bibfnamefont {M.}~\bibnamefont
  {Ablikim}} \emph {et~al.} (\bibinfo {collaboration} {BESIII}),\ }\href
  {https://doi.org/10.1007/JHEP05(2022)155} {\bibfield  {journal} {\bibinfo
  {journal} {JHEP}\ }\textbf {\bibinfo {volume} {05}},\ \bibinfo {pages}
  {155}},\ \Eprint {https://arxiv.org/abs/2112.06477} {arXiv:2112.06477
  [hep-ex]} \BibitemShut {NoStop}%
\bibitem [{\citenamefont {Zhukova}\ \emph {et~al.}(2018)\citenamefont {Zhukova}
  \emph {et~al.}}]{Belle:2017grj}%
  \BibitemOpen
  \bibfield  {author} {\bibinfo {author} {\bibfnamefont {V.}~\bibnamefont
  {Zhukova}} \emph {et~al.} (\bibinfo {collaboration} {Belle}),\ }\href
  {https://doi.org/10.1103/PhysRevD.97.012002} {\bibfield  {journal} {\bibinfo
  {journal} {Phys. Rev. D}\ }\textbf {\bibinfo {volume} {97}},\ \bibinfo
  {pages} {012002} (\bibinfo {year} {2018})},\ \Eprint
  {https://arxiv.org/abs/1707.09167} {arXiv:1707.09167 [hep-ex]} \BibitemShut
  {NoStop}%
\bibitem [{\citenamefont {Cronin-Hennessy}\ \emph {et~al.}(2009)\citenamefont
  {Cronin-Hennessy} \emph {et~al.}}]{CLEO:2008ojp}%
  \BibitemOpen
  \bibfield  {author} {\bibinfo {author} {\bibfnamefont {D.}~\bibnamefont
  {Cronin-Hennessy}} \emph {et~al.} (\bibinfo {collaboration} {CLEO
  Collaboration}),\ }\href {https://doi.org/10.1103/PhysRevD.80.072001}
  {\bibfield  {journal} {\bibinfo  {journal} {Phys.\ Rev.\ D}\ }\textbf
  {\bibinfo {volume} {{\bf 80}}},\ \bibinfo {pages} {072001} (\bibinfo {year}
  {2009})},\ \Eprint {https://arxiv.org/abs/0801.3418} {arXiv:0801.3418
  [hep-ex]} \BibitemShut {NoStop}%
\bibitem [{\citenamefont {Dong}\ \emph {et~al.}(2018)\citenamefont {Dong},
  \citenamefont {Wang},\ and\ \citenamefont {Yuan}}]{Dong:2017tpt}%
  \BibitemOpen
  \bibfield  {author} {\bibinfo {author} {\bibfnamefont {X.-K.}\ \bibnamefont
  {Dong}}, \bibinfo {author} {\bibfnamefont {L.-L.}\ \bibnamefont {Wang}},\
  and\ \bibinfo {author} {\bibfnamefont {C.-Z.}\ \bibnamefont {Yuan}},\ }\href
  {https://doi.org/10.1088/1674-1137/42/4/043002} {\bibfield  {journal}
  {\bibinfo  {journal} {Chin. Phys. C}\ }\textbf {\bibinfo {volume} {42}},\
  \bibinfo {pages} {043002} (\bibinfo {year} {2018})},\ \Eprint
  {https://arxiv.org/abs/1711.07311} {arXiv:1711.07311 [hep-ex]} \BibitemShut
  {NoStop}%
\bibitem [{\citenamefont {Schmidt}\ and\ \citenamefont
  {Steinhauser}(2012)}]{Schmidt:2012az}%
  \BibitemOpen
  \bibfield  {author} {\bibinfo {author} {\bibfnamefont {B.}~\bibnamefont
  {Schmidt}}\ and\ \bibinfo {author} {\bibfnamefont {M.}~\bibnamefont
  {Steinhauser}},\ }\href {https://doi.org/10.1016/j.cpc.2012.03.023}
  {\bibfield  {journal} {\bibinfo  {journal} {Comput.\ Phys.\ Commun.}\
  }\textbf {\bibinfo {volume} {{\bf 183}}},\ \bibinfo {pages} {1845} (\bibinfo
  {year} {2012})},\ \Eprint {https://arxiv.org/abs/1201.6149} {arXiv:1201.6149
  [hep-ph]} \BibitemShut {NoStop}%
\bibitem [{\citenamefont {Barnes}\ \emph {et~al.}(1997)\citenamefont {Barnes},
  \citenamefont {Close}, \citenamefont {Page},\ and\ \citenamefont
  {Swanson}}]{Barnes:1996ff}%
  \BibitemOpen
  \bibfield  {author} {\bibinfo {author} {\bibfnamefont {T.}~\bibnamefont
  {Barnes}}, \bibinfo {author} {\bibfnamefont {F.}~\bibnamefont {Close}},
  \bibinfo {author} {\bibfnamefont {P.}~\bibnamefont {Page}},\ and\ \bibinfo
  {author} {\bibfnamefont {E.}~\bibnamefont {Swanson}},\ }\href
  {https://doi.org/10.1103/PhysRevD.55.4157} {\bibfield  {journal} {\bibinfo
  {journal} {Phys.\ Rev.\ D}\ }\textbf {\bibinfo {volume} {{\bf 55}}},\
  \bibinfo {pages} {4157} (\bibinfo {year} {1997})},\ \Eprint
  {https://arxiv.org/abs/hep-ph/9609339} {arXiv:hep-ph/9609339} \BibitemShut
  {NoStop}%
\bibitem [{\citenamefont {Bai}\ \emph {et~al.}(2002)\citenamefont {Bai} \emph
  {et~al.}}]{BES:2001ckj}%
  \BibitemOpen
  \bibfield  {author} {\bibinfo {author} {\bibfnamefont {J.}~\bibnamefont
  {Bai}} \emph {et~al.} (\bibinfo {collaboration} {BES Collaboration}),\ }\href
  {https://doi.org/10.1103/PhysRevLett.88.101802} {\bibfield  {journal}
  {\bibinfo  {journal} {Phys. Rev. Lett.}\ }\textbf {\bibinfo {volume} {88}},\
  \bibinfo {pages} {101802} (\bibinfo {year} {2002})},\ \Eprint
  {https://arxiv.org/abs/hep-ex/0102003} {arXiv:hep-ex/0102003} \BibitemShut
  {NoStop}%
\bibitem [{\citenamefont {Ablikim}\ \emph {et~al.}(2006)\citenamefont {Ablikim}
  \emph {et~al.}}]{Ablikim:2006mb}%
  \BibitemOpen
  \bibfield  {author} {\bibinfo {author} {\bibfnamefont {M.}~\bibnamefont
  {Ablikim}} \emph {et~al.},\ }\href
  {https://doi.org/10.1103/PhysRevLett.97.262001} {\bibfield  {journal}
  {\bibinfo  {journal} {Phys.\ Rev.\ Lett.}\ }\textbf {\bibinfo {volume} {{\bf
  97}}},\ \bibinfo {pages} {262001} (\bibinfo {year} {2006})},\ \Eprint
  {https://arxiv.org/abs/hep-ex/0612054} {arXiv:hep-ex/0612054} \BibitemShut
  {NoStop}%
\bibitem [{\citenamefont {Rapidis}\ \emph {et~al.}(1977)\citenamefont {Rapidis}
  \emph {et~al.}}]{Rapidis:1977cv}%
  \BibitemOpen
  \bibfield  {author} {\bibinfo {author} {\bibfnamefont {P.}~\bibnamefont
  {Rapidis}} \emph {et~al.},\ }\href
  {https://doi.org/10.1103/PhysRevLett.39.526} {\bibfield  {journal} {\bibinfo
  {journal} {Phys.\ Rev.\ Lett.}\ }\textbf {\bibinfo {volume} {{\bf 39}}},\
  \bibinfo {pages} {526} (\bibinfo {year} {1977})},\ \bibinfo {note} {[Erratum:
  Phys.\ Rev.\ Lett.\ {\bf 39}, 974 (1977)]}\BibitemShut {NoStop}%
\bibitem [{\citenamefont {Osterheld}\ \emph {et~al.}(1986)\citenamefont
  {Osterheld} \emph {et~al.}}]{Osterheld:1986hw}%
  \BibitemOpen
  \bibfield  {author} {\bibinfo {author} {\bibfnamefont {A.}~\bibnamefont
  {Osterheld}} \emph {et~al.},\ }\href@noop {} {\bibinfo {title} {{\em
  Measurements of Total Hadronic and Inclusive $D^*$ Cross-Sections in $e^+
  e^-$ Annihilations Between 3.87~{GeV} and 4.5~{GeV}}}} (\bibinfo {year}
  {1986})\BibitemShut {NoStop}%
\bibitem [{\citenamefont {Schindler}(1979)}]{Schindler:1979rj}%
  \BibitemOpen
  \bibfield  {author} {\bibinfo {author} {\bibfnamefont {R.}~\bibnamefont
  {Schindler}},\ }\emph {\bibinfo {title} {{\it Charmed Meson Production and
  Decay Properties at the $\psi^{\prime\prime}$(3770)}}},\ \href@noop {}
  {\bibinfo {type} {master's thesis}} (\bibinfo {year} {1979})\BibitemShut
  {NoStop}%
\end{thebibliography}%

\end{document}